\documentclass[fleqn,usenatbib]{mnras}

\usepackage{newtxtext,newtxmath}
\usepackage[T1]{fontenc}

\DeclareRobustCommand{\VAN}[3]{#2}
\let\VANthebibliography\thebibliography
\def\thebibliography{\DeclareRobustCommand{\VAN}[3]{##3}\VANthebibliography}

\usepackage{graphicx}	% Including figure files
\usepackage{amsmath}	% Advanced maths commands
\usepackage{color}
\usepackage{enumitem}
\usepackage{hyperref}
\usepackage{verbatim}
\usepackage{mathrsfs}
\usepackage[all]{hypcap} 
\usepackage{afterpage}    
\usepackage{marginnote}
\usepackage{multirow}
\usepackage{lineno}
\usepackage{float}
\usepackage{siunitx}
\title[Decomposing Magnetic Fields in the CMZ]{Decomposing Magnetic Fields in Three Dimensions over the Central Molecular Zone}

\author[Hu, Lazarian \& Wang]{
Yue Hu$^{1,2}$\thanks{E-mail: yue.hu@wisc.edu}
,A. Lazarian$^{2,3}$\thanks{E-mail: alazarian@facstaff.wisc.edu}
,Q. Daniel Wang$^{4}$\thanks{E-mail: wqd@astro.umass.edu}
\\
$^{1}$Department of Physics, University of Wisconsin-Madison, Madison, WI, 53706, USA\\
$^{2}$Department of Astronomy, University of Wisconsin-Madison, Madison, WI, 53706, USA\\
$^{3}$Centro de Investigación en Astronomía, Universidad Bernardo O’Higgins, Santiago, General Gana 1760, 8370993,
Chile\\
$^{4}$Department of Astronomy, University of Massachusetts, Amherst, MA 01003, USA\\
}

\date{Accepted XXX. Received YYY; in original form ZZZ}

\pubyear{2021}

\begin{document}
\label{firstpage}
\pagerange{\pageref{firstpage}--\pageref{lastpage}}
\maketitle

\begin{abstract}
Measuring magnetic fields in the interstellar medium and obtaining their distribution along line-of-sight is very challenging with the traditional techniques. The Velocity Gradient Technique (VGT), which utilizes anisotropy of magnetohydrodynamic (MHD) turbulence, provides an attractive solution. Targeting the central molecular zone (CMZ), we test this approach by applying the VGT to $\rm ^{12}CO$ and $\rm ^{13}CO$ (J = 1-0) data cubes. We first used the SCOUSEPY algorithm to decompose the CO line emissions into separate velocity components, and then we constructed pseudo-Stokes parameters via the VGT to map the plane-of-the-sky magnetic fields in three-dimension. We present the decomposed magnetic field maps and investigate their significance. While the line-of-sight integrated magnetic field orientation is shown to be consistent with the polarized dust emission from the Planck survey at 353 GHz, individual velocity components may exhibit different magnetic fields. We present a scheme of magnetic field configuration in the CMZ based on the decomposed magnetic fields. In particular, we observe a nearly vertical magnetic field orientation in the dense clump near the Sgr B2 and a change in the outflow regions around the Sgr A*. Two high-velocity structures associated with an expanding ring in the CMZ show distinct swirling magnetic field structures. These results demonstrate the potential power of the VGT to decompose velocity or density-dependent magnetic structures. 
\end{abstract}


%%%%%%%%%%%%%%%%%%%%%%%%%%%%%%%%%%%%%%%%%%%%%%%%%%

%%%%%%%%%%%%%%%%% BODY OF PAPER %%%%%%%%%%%%%%%%%%

\section{Introduction}
Magnetic fields are expected to be crucial in regulating the structure and evolution of the circumnuclear medium of galaxies \citep{2009A&A...505.1183F,2010Natur.463...65C,2014MNRAS.440.3370K,2019A&A...630A..74M}. They play a central role in such physical processes as cosmic ray propagation and star formation \citep{2012ApJ...761..156F,2014PhR...539...49K,2018ApJ...863..118B,Hu20,2022arXiv220301508H}. Particularly relevant is the unexpectedly lower star formation rate observed in the central molecular zone (CMZ) of our Galaxy \citep{2013MNRAS.429..987L,2017MNRAS.469.2263B}. However, despite its apparent importance, the magnetic field in the CMZ remains poorly understood.

Several approaches to measure the magnetic field in the CMZ have been proposed. For instance, two of the most typical measurements are observations of polarized dust emission or synchrotron radio observation \citep{1988ApJ...333..729W,1990ApJ...362..114H,2000ApJ...529..241N,2005MNRAS.360.1305R,2008A&A...478..435R,2007SPIE.6678E..0DV}. The former relies on the fact that the dust grain's semi-minor axis aligns with ambient magnetic field lines as driven mainly by radiative torques \citep{2015ARA&A..53..501A,2007ApJ...669L..77L}. The recent Planck survey of polarized dust emission at a wavelength of \SI{850}{\micro\meter} and \SI{53}{\micro\meter} HAWC+ survey have considerably advanced our knowledge of the magnetic field orientations in the CMZ \citep{2015A&A...576A.104P,2018JAI.....740008H}. In addition, the polarized radio observation further completes the picture of magnetic fields at typically ionized gas phase \citep{1986AJ.....92..818T,1990IAUS..140..361M,1999ApJ...521L..41L,2019ApJ...884..170P}. However, these measurements are inherently limited to only the line-of-sight (LOS) projected averages of magnetic properties or a single component of the three-dimensional magnetic field.

The Velocity Gradient Technique (VGT; \citealt{GL17,YL17a, LY18a, PCA}) provides a solution to disentangle the projected plane-of-the-sky (POS) magnetic fields along the LOS. Unlike the synchrotron or dust polarization, VGT derives the orientation of the magnetic field from anisotropic magnetohydrodynamic (MHD) turbulence. Because turbulent eddy is statistically elongating along the local magnetic fields that percolate them \citep{GS95,LV99}, the gradient of velocity fluctuations' amplitude is perpendicular to the local magnetic field. Consequently, the velocity gradient rotated by 90$^\circ$ reveals the magnetic field orientation. 

The VGT has been applied to trace the integrated magnetic field in the CMZ \citep{CMZ}. Multiple-wavelength observations, including spectroscopic and radio data, were employed to disentangle the magnetic fields embedded in different gas phases along the LOS. The measurements allowed us to map the wavelength-dependent magnetic field in diffuse-ionized-gas and dense-cold-gas clouds on multi scales from 10 pc to 0.1 pc over the entire CMZ. In general, the magnetic fields traced by the VGT were globally compatible with the polarization measurements, accounting for the contribution from the galactic foreground/background. This successful application encouraged us to investigate further the VGT's applicability in decomposing the projected POS magnetic fields along the LOS. The decomposition is performed in PPV space using emission lines. The velocity information there advantageously reveals the 3D spatial position in real space. For instance, the galactic disk contribution is largely concentrated at velocity $~0$ km/s, while molecular gas observed at high velocity $\pm200$ km/s is associated with an expanding ring \citep{1972ApJ...175L.127S,1974PASJ...26..117K,2019PASJ...71S..19T}. Therefore, the decomposed magnetic fields probed by the VGT provide a new view of the 3D magnetic field configuration in the CMZ region.

The decomposition is achieved by making a synergy of the VGT, the Python implementation of the Semi-Automated multi-COmponent Universal Spectral-line fitting Engine (SCOUSEPY; \citealt{2016MNRAS.457.2675H}) \footnote{\url{https://github.com/jdh
enshaw/scousepy}}, and the Agglomerative Clustering for ORganising Nested Structures (ACORNS) \footnote{\url{https://github.com/jdhen
shaw/acorns.}}. SCOUSEPY is a routine used to fit large quantities of complex spectroscopic data efficiently and systematically and ACORNS is a hierarchical agglomerative clustering technique. The SCOUSEPY decomposes spectroscopic emission lines into various Gaussian components, which are then grouped via the ACORNS. Application of the VGT to the grouped components results in a velocity-resolved mapping of the magnetic fields. We test this approach on $\rm ^{12}CO$ (1–0) and $\rm ^{13}CO$ (1–0) emission lines and compare the VGT mapping of the magnetic fields with the Planck 353 GHz polarized dust emission.

%In order to have a more comprehensive picture of the magnetic fields, our analysis includes the magnetic fields in neighboring physically different regions. For instance, the Radio Arc, the Arched Filaments, and Sagittarius A West \citep{2010ApJS..191..275L,2012ApJ...755...90I}. The first two structures observed at 1.4 GHz suggest either poloidal or toroidal magnetic fields in the CMZ. While the Sgr A West is observed with [\ion{Ne}{2}] emission and it has a very distinctive and extreme physical condition due to the central supermassive black hole. These data, including the molecular lines, the Planck 353 GHz \citep{2020A&A...641A...3P} and HWAC+ 53$\mu m$ \citep{2018JAI.....740008H} dust polarization data, allow us to map  the wavelength-dependent magnetic field in both diffuse-ionized-gas region and dense-cold-gas region in multi scales from the order of 10 pc to 0.1 pc. Thus, this work provides a multi-wavelength and multi-scale view of the magnetic fields and their interactions with the multi-phase gas in different parts of the galactic center ecosystem. 

The paper is organized as follows. In \S~\ref{sec:data}, we provide the details of the observational data used in this work. In \S~\ref{sec:method}, we illustrate the methodology of VGT to trace the magnetic fields. In \S~\ref{sec:result}, we present the magnetic fields traced by VGT and compare the magnetic fields traced by VGT with Planck 353 GHz polarization. We decompose emission lines into several velocity components and derive their magnetic field orientation. We give discussion in \S~\ref{sec:dis} and summary in \S~\ref{sec:con}, respectively.

\section{Observational Data}
\label{sec:data}
\subsection{CO emission lines}
The $\rm ^{12}CO$ (1–0) and $\rm ^{13}CO$ (1–0) emission lines towards the Galactic CMZ were observed with the Nobeyama 45m telescope \citep{2019PASJ...71S..19T}. The data cover the area: $-0.8^\circ <l<1.2^\circ$ and $-0.35^\circ <b<+0.35^\circ$ with a beamwidth of $\rm FWHM\approx15''$ and velocity resolution of $\approx1.3$ km s$^{-1}$ for $\rm ^{12}CO$ (1–0) and $\approx0.67$ km s$^{-1}$ for $\rm ^{13}CO$ (1–0). The final data cubes were resampled onto a $7.5'' \times 7.5''\times 2$ km s$^{-1}$ grid resolution. The rms noises of $\rm ^{12}CO$ (1–0) and $\rm ^{13}CO$ (1–0) are approximately 1.00 K and 0.20 K, respectively. Details of the data reduction can be found in  \citep{2019PASJ...71S..19T}. We select the emission within the radial velocity range of -220 to +220 km s$^{-1}$ for our analysis.

\subsection{Dust polarization}
To obtain the magnetic field orientation from polarized dust emission, we adopt the data from Planck 3rd Public Data Release \citep{2020A&A...641A..12P} for this work. The magnetic field orientation $\phi_B=\phi+\pi/2$ is inferred from the polarization angle $\phi$:
\begin{equation}
\begin{aligned}
    \phi&=\frac{1}{2}\arctan(-U,Q)\\
%p&=\sqrt{Q^2+U^2}/I
\end{aligned}
\end{equation}
where $Q$ and $U$ are Stokes parameters of polarized dust emission. $-U$ resolves the angle conversion from the HEALPix system to the IAU system. We minimize the noise on the maps by smoothing the results with a Gaussian filter FWHM$\sim10'$, which is twice larger than Planck's beam size $\sim5'$. 

In order to remove the signal from polarization galactic foreground and background, we adopt the same method as \citet{CMZ}. By assuming that emission in these referenced regions provides spatial uniformity foreground emission, we calculate the average Stokes $I$, $Q$, and $U$ in that region, and the mean values were then subtracted from each of the $I$, $Q$, $U$ map of Planck polarization. To do so, we select two regions as the foreground reference point. One region is located at the west of the CMZ, spanning from $l = 3^\circ$ to $18^\circ$ and $b = -0.35^\circ$ to $0.35^\circ$ and another one is located at the east of the CMZ spanning from $l = -18^\circ$ to $-3^\circ$ and $b = -0.35^\circ$ to $0.35^\circ$ (see \citealt{CMZ} for details).

\section{Methodology}
\label{sec:method}
\subsection{Theory of MHD turbulence}
The MHD turbulence theory (\citealt{GS95}, noted as GS95 hereafter) and the fast turbulent reconnection theory (\citealt{LV99}, noted as LV99 hereafter) form the theoretical foundations of VGT. GS95 proposed the picture of anisotropic MHD turbulence considering the "critical balance" condition, in which the cascading time ($k_{\perp}v_l)^{-1}$ is equal to the wave period ($k_{\parallel}v_A)^{-1}$. Here $k_{\perp}$ and $k_{\parallel}$ donate wavevectors perpendicular and parallel to the magnetic field direction, respectively, while $v_l$ is the turbulent velocity at scale $l$ and $v_A$ is the Alfv\'en speed. Within this condition, Kolmogorov-type turbulence, which obeys $v_l \propto l^{1/3}$, exhibit an anisotropic scaling in Fourier space:
\begin{equation}
\begin{aligned}
          k_{\parallel} \propto k_{\perp}^{2/3}
\end{aligned}
\label{aligned}
\end{equation}
This scaling relation suggests that MHD turbulent eddies are elongating along magnetic fields exhibiting anisotropy. This anisotropy is not observed in the Fourier space and the reason is that the critical balance should be formulated in the local frame of the turbulent motions considered, as we discuss below. At the same time, in Fourier space, the local spatial information is not available so that the anisotropy is measured with respect to the mean magnetic field, which builds up the global reference frame. In this frame, anisotropy of larger eddies, which is most significant, dominates over small eddies \citep{2000ApJ...539..273C}. Therefore, the anisotropy in the global reference frame is scale-independent \citep{2000ApJ...539..273C,2002ApJ...564..291C,2021ApJ...911...37H}, despite the turbulent fluctuation alone being scale-dependent.

The anisotropy scaling in the local reference frame, defined by the local magnetic field passing through the eddy at scale $l$, is later derived by LV99. The difference between the local cascade and that given by Eq.~(\ref{aligned}) is that instead of the mean magnetic field, the motions are considered with respect to the local magnetic field at scale $l$. \footnote{It is intuitively clear that this is the only right approach to MHD turbulence. The motions on a small scale $l$ cannot be affected by magnetic field on a much larger or much smaller scales. Therefore the direction of the magnetic field surrounding the eddy is important for defining the parallel and perpendicular scales that enter the critical balance. The importance of the local system of reference is essential for understanding of MHD turbulence.}

Consequently, turbulence cascade is channeled by eddies with axes aligned with the local magnetic field direction. Such eddies, due to the fast magnetic reconnection (see LV99) mix up magnetic field in perpendicular direction in the hydrodynamic-type manner. Thus, the MHD cascade obeys the hydrodynamic-type Kolmogorov law $v_{l,\bot} \propto l^{1/3}_\bot$ in diffuse ISM \citep{1995ApJ...443..209A,2010ApJ...710..853C}\footnote{In highly compressive ISM, the scaling can be different from Kolmogorov law, but the MHD turbulence is still anisotropic and this does not affect the direction of velocity gradient \citep{LY18a}.}, where $v_{l,\bot}$ is the turbulent velocity perpendicular to the local magnetic field at scale $l$. The anisotropy scaling in the local reference frame is given in LV99:
\begin{equation}
\begin{aligned}
l_{\parallel}  =  L_{\rm inj}(\frac{l_{\bot}}{L_{\rm inj}})^{2/3}M_{\rm A}^{-4/3}, M_{\rm A} \le 1
\end{aligned}
\end{equation}
where $l_{\parallel}$ denotes the scale parallel to the local magnetic field, $L_{\rm inj}$ is the turbulent injection scale, and the Alfv\'en Mach number is $M_{\rm A}$. The scale-dependent anisotropy is only observable in this local reference frame. Furthermore, the anisotropy scaling for velocity fluctuation is given as (LV99):
\begin{equation}
\begin{aligned}
    &v_{l,\bot} = v_{\rm inj}(\frac{l_{\bot}}{L_{\rm inj}})^{1/3}M_{\rm A}^{1/3}\\
\end{aligned}
\end{equation}
where $v_{\rm inj}$ corresponds to the injection velocity. The anisotropy means that the maximum fluctuation appears in the direction perpendicular to the magnetic field \citep{2021ApJ...911...37H,2021ApJ...915...67H}. The gradient of velocity fluctuation is therefore along the perpendicular direction and it scales as:
\begin{equation}
\begin{aligned}
    &\nabla v_l\propto\frac{v_{l,\bot}}{l_{\bot}}\approx\frac{v_{\rm inj}}{L_{\rm inj}}M_{\rm A}^{1/3}(\frac{l_{\bot}}{L_{\rm inj}})^{-2/3}
\end{aligned}
\end{equation}
The amplitude of the velocity gradient is also scale-dependent and is more significant at small scales. This property ensures that at small scales, other effects such as galactic shear, have insignificant contribution to the velocity gradient (see \citealt{CMZ}). Note that only the local reference frame maintains spatial information allowing tracing of the local magnetic field via the velocity gradient.

Gradients and the effects of noise on gradients were studied empirically in \cite{2011Natur.478..214G} and \cite{2012ApJ...749..145B} for gradients of the Stokes parameters (see also \citealt{2015MNRAS.451..372R} and \citealt{2017MNRAS.468.2957R}). The ability of the velocity caustics to trace the magnetic field were demonstrated in \citet{LY18a}. There it was demonstrated that the removal of low spatial frequencies from the data does not negatively affect the tracing of the restored POS magnetic field structure. Theoretically, this was proven in \citet{2020arXiv200207996L} and demonstrated numerically and observationally in \citet{kspace}. This gives us confidence that the gradient-obtained magnetic fields represent the actual POS magnetic field structure. 

\begin{figure*}
	\includegraphics[width=1\linewidth]{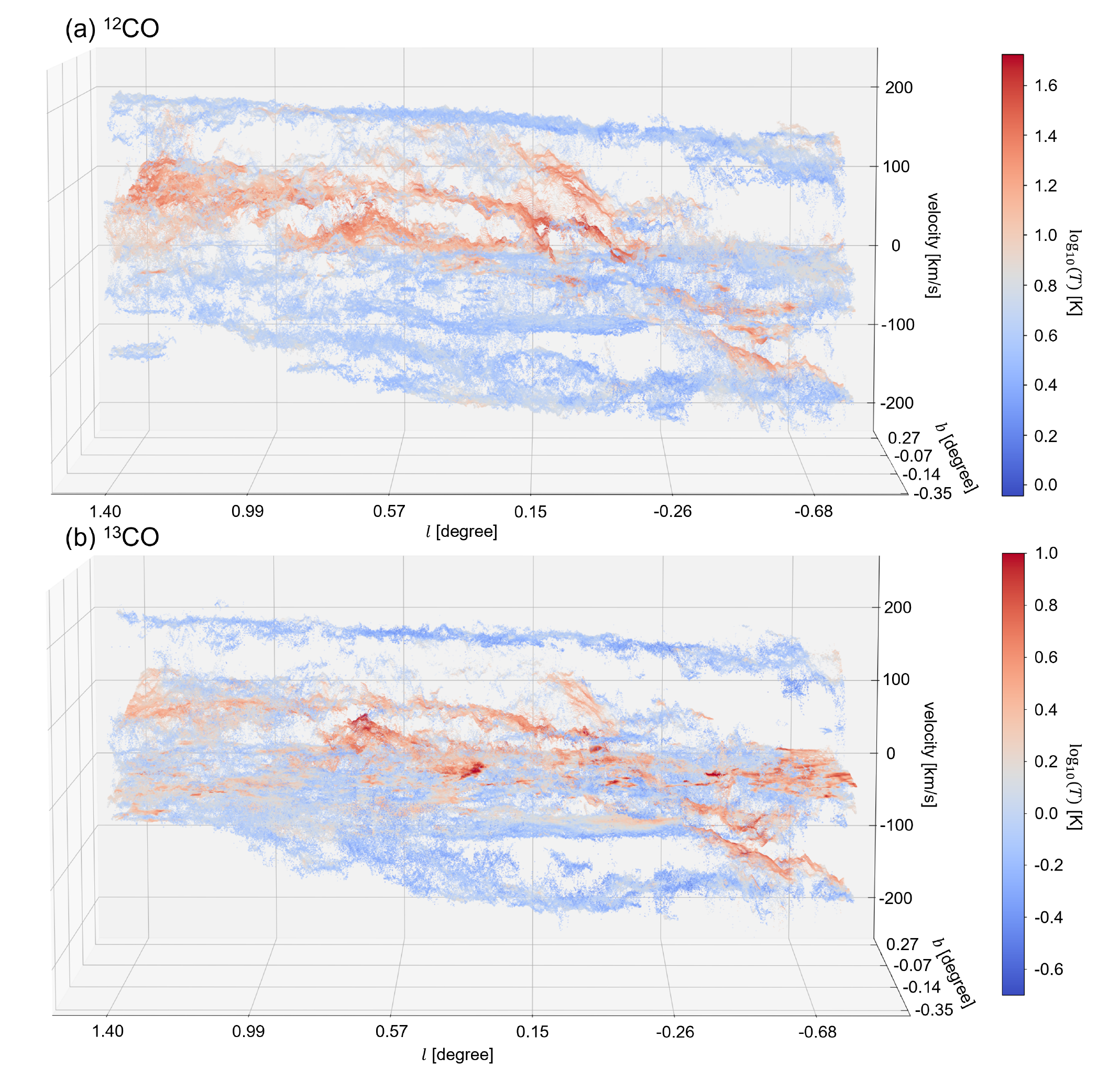}
    \caption{Visualization of the decomposed $\rm ^{12}CO$ (top) and $\rm ^{13}CO$ (bottom). Color-coded according to the emission line intensity peaks.}
    \label{fig:1}
\end{figure*}

\begin{figure*}
	\includegraphics[width=1\linewidth]{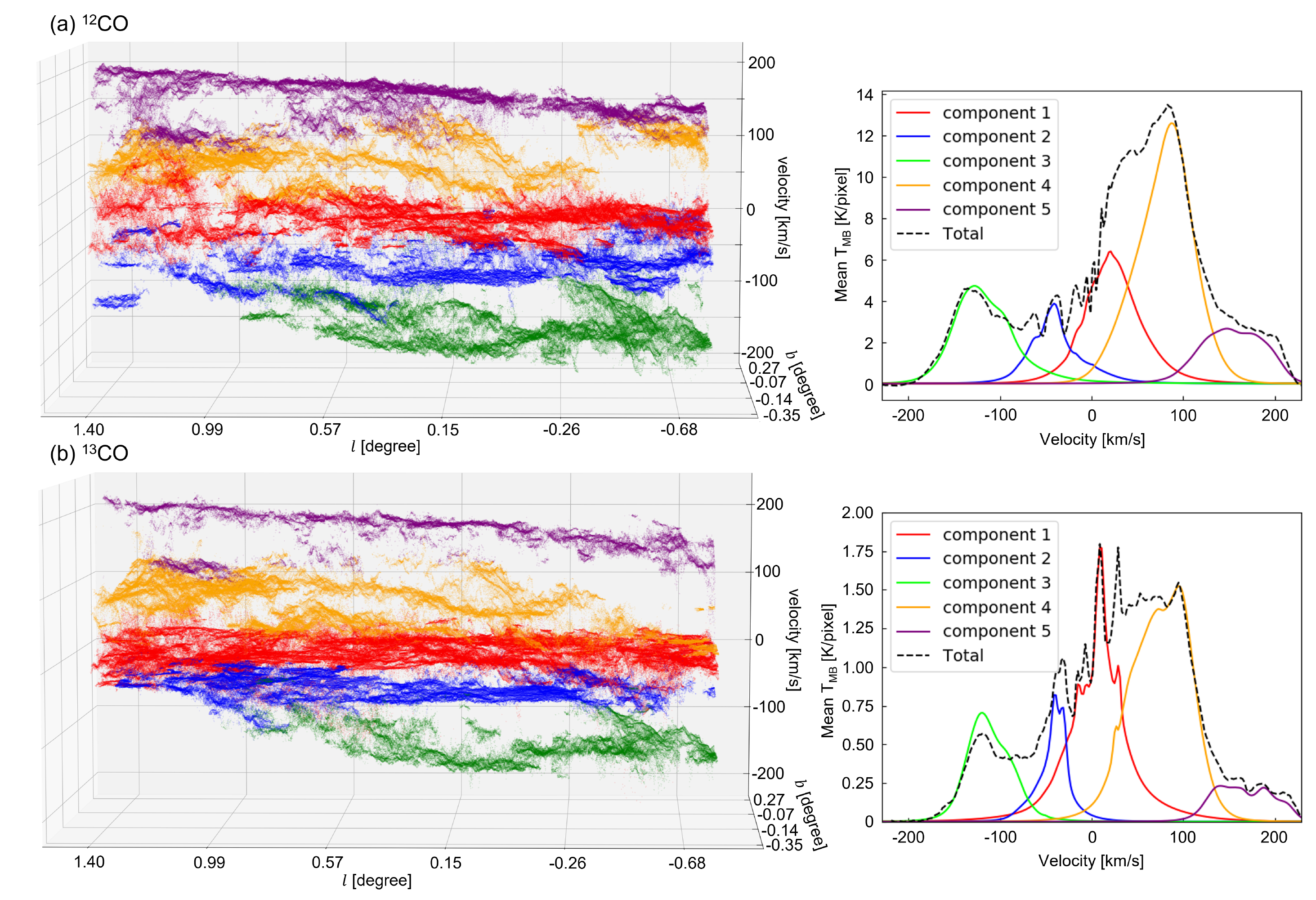}
    \caption{Visualization of the decomposed emission lines $\rm ^{12}CO$ (top left) and $\rm ^{13}CO$ (bottom left), as well as the averaged spectrum (right). Colors are used to distinguish the clustered five components.}
    \label{fig:2}
\end{figure*}

\begin{figure*}
	\includegraphics[width=1\linewidth]{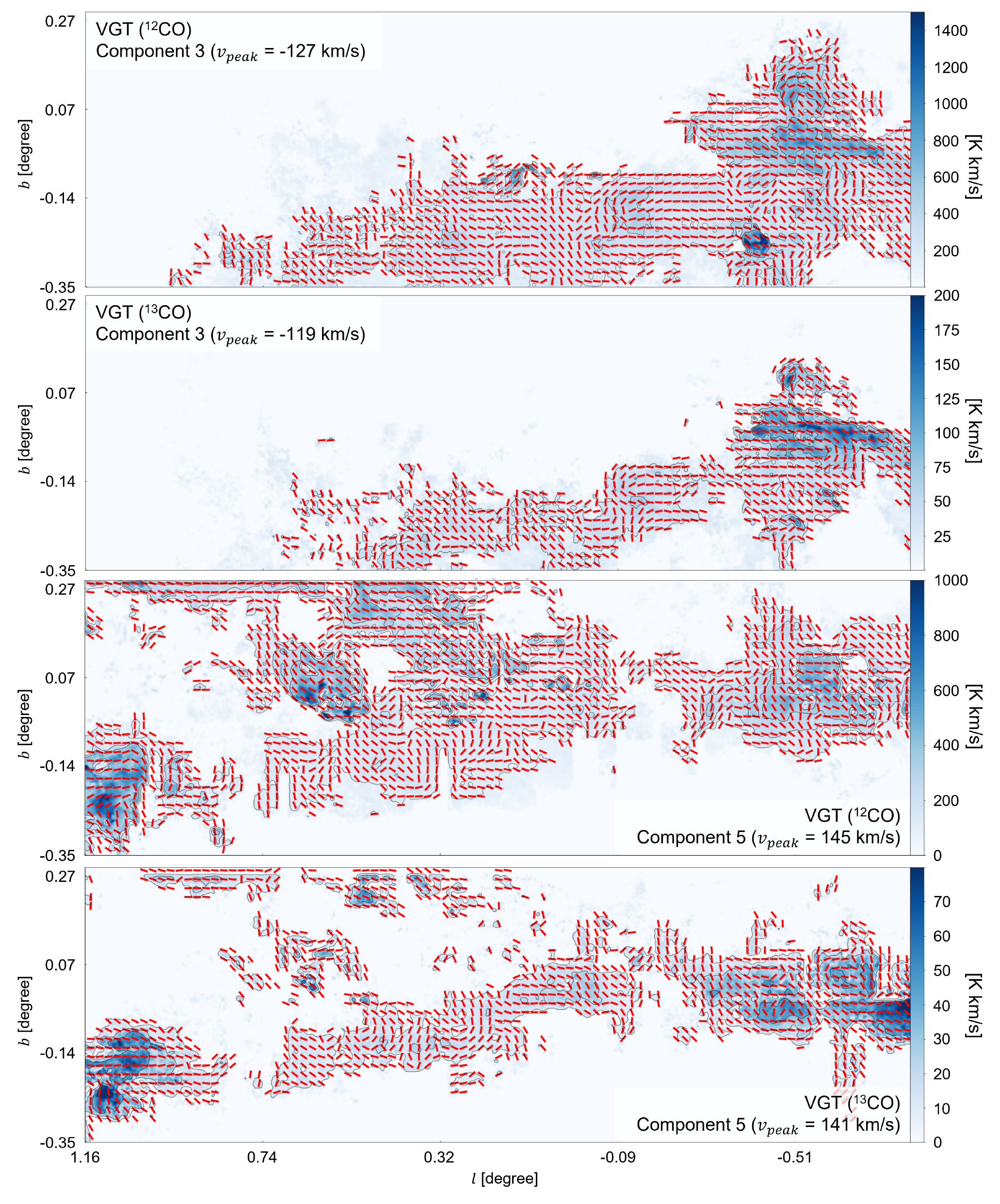}
    \caption{Visualization of the magnetic field (red segment) associated with the decomposed velocity component 3 and 5 (see Fig.~\ref{fig:2}). The magnetic field is obtained from the VGT using $\rm ^{12}CO$ and $\rm ^{13}CO$.}
    \label{fig:3}
\end{figure*}

\subsection{The Velocity Gradient Technique}
The VGT can employ either a velocity centroid map or a thin velocity channel map to extract statistics of velocity fluctuation. The latter is rooted in the non-linear spectroscopic mapping of Doppler-shifted line, which is also called velocity caustics effect analytically explored in \cite{2000ApJ...537..720L}. It was shown that the contribution of velocity fluctuations to the observed intensity fluctuation in Position-Position-Velocity (PPV) space increases compared to the contributions of density fluctuation when the channel width is sufficiently thin. Specifically, when a channel is sufficiently thin satisfying the criterion:
\begin{equation}
\begin{aligned}
        &\Delta v < \sqrt{\delta (v^2)}, \text{thin channel}\\
         &\Delta v \geq \sqrt{\delta (v^2)}, \text{thick channel}
\end{aligned}
\end{equation}
the velocity fluctuations get dominated. Here $\Delta v$ is the channel width and $\sqrt{\delta (v^2)}$ represents velocity dispersion corresponding to eddies under study. Accordingly, the gradient calculated from thin velocity channel is called Velocity Channel Gradients (VChGs; \citealt{LY18a}). The validity of velocity caustics effect in the multiple-phase medium of neutral hydrogen was questioned by \citet{2019ApJ...874..171C}. However, the problems with these sorts of arguments were exposed by \citet{2019arXiv190403173Y} with the analysis of observational data in \citet{2021ApJ...910..161Y}. It demonstrates the importance of velocity caustics in the multi-phase galactic H I.

In this work, we use VChGs to trace magnetic fields. The recipe of VChGs is presented in the following. Given a thin velocity channel \textbf{Ch(x,y)}, its x, y gradient components $\bigtriangledown_x {\rm Ch}_i (x,y)$ and $\bigtriangledown_y {\rm Ch}_i (x,y)$ are:
\begin{equation}
        \begin{aligned}
           \bigtriangledown_x \textit{${\rm Ch}_i (x,y)$}  =  \textit{$G_x$} * \textit{${\rm Ch}_i (x,y)$} \\
           \bigtriangledown_y \textit{${\rm Ch}_i (x,y)$}  =  \textit{$G_y$} * \textit{${\rm Ch}_i (x,y)$}
        \end{aligned}
        \end{equation}
where $G_x$, $G_y$ are Sobel kernels
$$
	G_x=\begin{pmatrix} 
	-1 & 0 & +1 \\
	-2 & 0 & +2 \\
	-1 & 0 & +1
	\end{pmatrix} \quad,\quad
	G_y=\begin{pmatrix} 
	-1 & -2 & -1 \\
	0 & 0 & 0 \\
	+1 & +2 & +1
	\end{pmatrix}
$$
and $*$ denotes convolution. The pixelized gradient map $\psi_g^i (x,y)$ is then:
\begin{equation}
       \begin{aligned}
%                   &G_x = 
%        \begin{pmatrix}
%         -1 & 0 & +1\\ 
%         -2 & 0 & +2\\
%         -1 & 0 & +1
%        \end{pmatrix},
%        G_y =
%        \begin{pmatrix}
%         -1 & -2 & +1\\ 
%         0 & 0 & 0\\
%         +1 & +2 & +1
%        \end{pmatrix}\\
&\psi^i_g (x,y)=\tan^{-1}(\frac{\bigtriangledown_y {\rm Ch}_i (x,y)}{\bigtriangledown_x {\rm Ch}_i (x,y)})
\end{aligned}
\end{equation}

Another critical step is averaging the gradient within a sub-region of interests as introduced by \cite{YL17a}. The sub-block averaging sets sub-blocks in the gradient maps and applies a Gaussian fitting method to the distribution of the gradient's orientation. The peak value of the Gaussian distribution gives the most probable orientation of the gradients. We average the gradients over each 20$\times$20 pixels sub-block, which is an empirically and numerically minimum block size \citep{H2}. The dispersion of gradients within a sub-block depends on the medium's magnetization level \citep{2018ApJ...865...46L}. 

For each thin velocity channel, this averaging method is applied. We denoted the gradient map after averaging as $\psi_{g,s}^i (x,y)$. Consequently, we acquire compiled gradient maps $\psi_{g,s}^i (x,y)$ of the same number as the thin velocity channel ($n_v$). The pseudo Stokes parameters, similar to the Stokes parameters used for polarization data, are constructed from:
\begin{equation}
        \begin{aligned}
           &Q_g (x,y)  =  \sum^{n_v}_{i=1} {\rm Ch}_i (x,y) \cos(2\psi^i_{g,s} (x,y))\\
           &U_g (x,y)  =  \sum^{n_v}_{i=1} {\rm Ch}_i (x,y) \sin(2\psi^i_{g,s} (x,y))\\
           &\psi_g = \frac{1}{2}\tan^{-1}(\frac{U_g}{Q_g})
        \end{aligned}
        \end{equation}
where $\psi_g$ donates the pseudo polarization angle, and thus the POS magnetic field direction is defined as $\psi_B = \psi_g + \pi/2$. In particular, pixels where the brightness temperature is less than three times the RMS noise level are blanked out. We smooth the gradients map $\psi_g$ with a Gaussian filter (FWHM $\sim1'$), which gives a higher resolution than \citet{CMZ}. The systematic uncertainty of the orientation accumulated all contribution across the velocity channels is typically less than $15^\circ$ (see the Appendix~A in \citealt{CMZ}). The uncertainty of decomposed single components could be lower.

The relative orientation between the two magnetic field directions probed with the Planck polarization ($\phi_B$) and the gradients ($\psi_B$) is measured with the Alignment Measure (AM; \citealt{GL17}), defined as
\begin{equation}
        \begin{aligned}
           {\rm AM}  =  2(\langle\cos^2(\theta_r)\rangle-\frac{1}{2})
        \end{aligned}
\end{equation}
where we have $\theta_r = \vert\phi_B-\psi_B\vert$ and $\langle...\rangle$ denote averaging over a region of interests. AM is a relative scale ranging from -1 to 1, with AM = 1 indicating that $\phi_B$ and $\psi_B$ being globally parallel, and AM = -1 denoting that the two are globally orthogonal. Note that the VGT has a higher resolution (i.e., smaller scale ) than the Planck measurement. The difference in resolutions may introduce uncertainty to the AM.

\subsection{Employing SCOUSEPY software}
To explore the VGT's potential in probing three-dimensional POS magnetic fields, we use the SCOUSEPY, a Python software based on the Semi-automated multi-COmponent Universal Spectral-line fitting Engine (SCOUSE) proposed by \cite{2016MNRAS.457.2675H,2019MNRAS.485.2457H}, to decompose each CO spectrum. SCOUSEPY is designed for fitting a large number of spectroscopic data with a multi-stages procedure. The software firstly divides the spatial data of the user-selected region into small areas, named Spectral Averaging Areas (SAAs), and outputs a spectrum averaged spatially for each of the SAAs. By manually marking the applicable velocity range of the data, then the fitting procedures is performed to find a Gaussian profile. More details can be found in \cite{2016MNRAS.457.2675H}.

In this work, we follow the tolerance criteria proposed in \cite{2016MNRAS.457.2675H}: (i) all detected components must have a brightness temperature greater than three times the local noise level; (ii) each Gaussian component must have an FWHM line-width of at least one channel: (iii) for two Gaussian components to be considered distinguishable, they must be separated by at least half of the FWHM of the narrowest of the two. (iv) the size of SAAs is set to ten pixels. Moreover, the decomposed spectrum is further processed by the ACORNS. We perform the ACORNS clustering only on the most robust spectral velocity components decomposed by SCOUSEPY, i.e. (i) the signal-to-noise ratio is more significant than three; (ii) the minimum radius of a cluster to be 20$''$, which is 130\% of the beam resolution; (iii) for two data points to be classified as "linked," the absolute difference in measured velocity dispersion can be no greater 10.0 km/s. The clustering comprises $\sim98\%$ of all data and results in several hierarchical and non-hierarchical components. Because both components contribute to VGT measurement, we combine them into the five most significant velocity components based on the un-decomposed averaged spectral lines. Note that the decomposition is performed for $\rm ^{12}CO$ and $\rm ^{13}CO$ separately, for which the clustering may not result in physically related components. To minimize this effect, firstly, we adopt the same tolerance criteria for both. Then we ensure the combined velocity component falls into a similar velocity range. Therefore, we expect the decomposed velocity components of $\rm ^{12}CO$ and $\rm ^{13}CO$ largely represent to the same physical structure, although a prospective joint decomposition could reduce the uncertainty.

\begin{figure*}
	\includegraphics[width=1.0\linewidth]{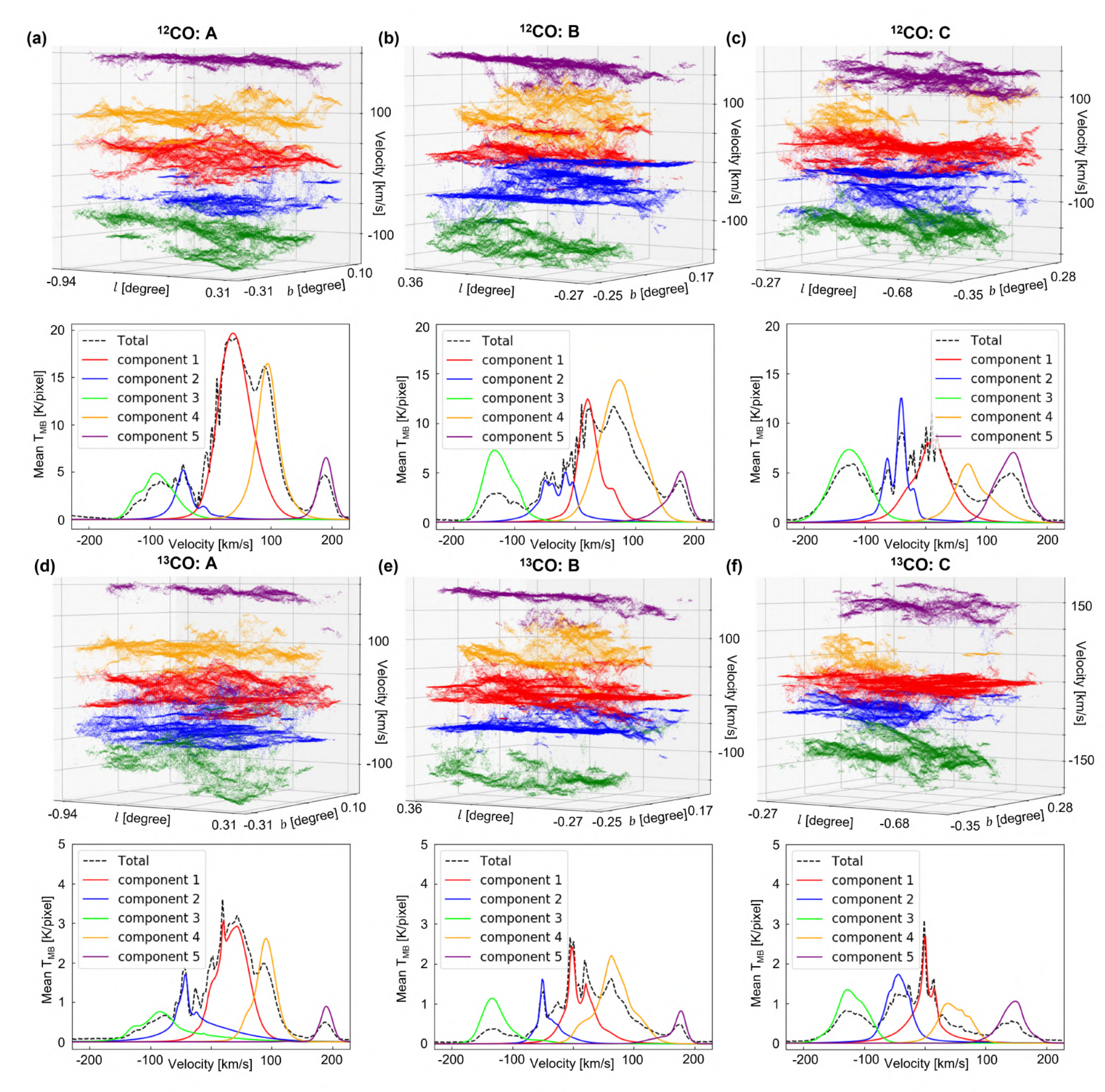}
    \caption{\textbf{Panel a:} visualization of the decomposed $\rm ^{12}CO$ spectrum in PPV space (top) and averaged $\rm ^{12}CO$ spectrum (bottom). The decomposition is performed for three sub-regions, i.e., \textbf{A} (left), \textbf{B} (middle), and \textbf{C} (right). \textbf{Panel b:} visualization of the decomposed $\rm ^{13}CO$ spectrum in PPV space (top) and averaged $\rm ^{13}CO$ spectrum (bottom). The decomposition is performed for the same three sub-regions. The dashed line represents un-decomposed averaged spectrum. }
    \label{fig:4}
\end{figure*}

\begin{figure*}
	\includegraphics[width=1\linewidth]{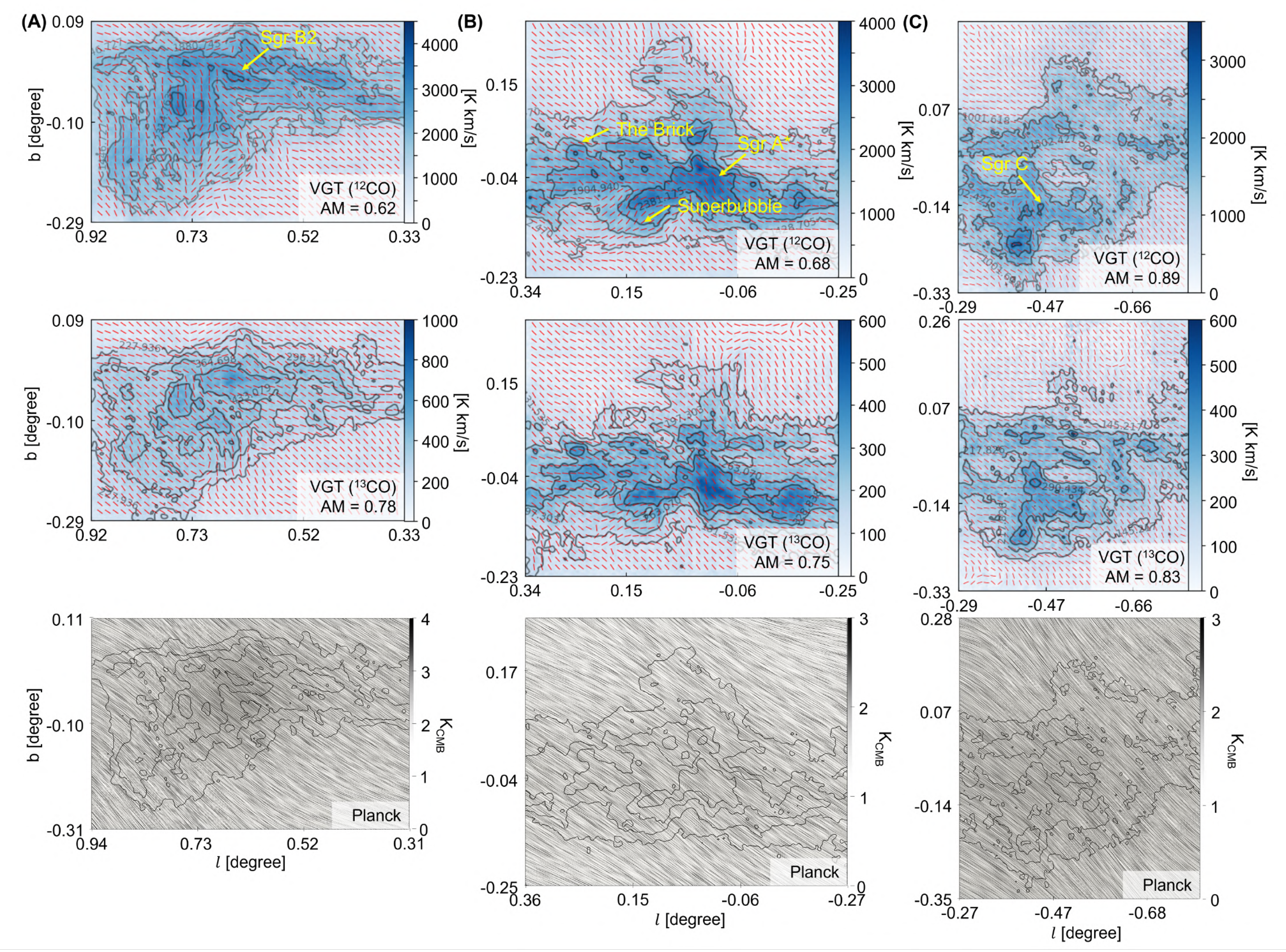}
    \caption{\textbf{Top \& Middle:} Visualization of magnetic fields (red segment) inferred from VGT using $\rm ^{12}CO$ and $\rm ^{13}CO$ emissions towards the CMZ. The CMZ is separated into three sub-regions, which are labeled as \textbf{A} (left), \textbf{B} (middle), and \textbf{C} (right), respectively. The magnetic fields are overlaid on corresponding integrated intensity color maps. The contours on $\rm ^{12}CO$ and $\rm ^{13}CO$ intensity maps start from the median intensity. \textbf{Bottom:} Visualization of magnetic fields inferred from Planck's polarized dust emission at 353 GHz towards the CMZ. The magnetic field is visualized using the Line Integral Convolution (LIC). The magnetic fields are overlaid on corresponding intensity color maps of polarized dust emission. The $\rm ^{12}CO$'s contours are also overlaid on the Planck maps.}
    \label{fig:5}
\end{figure*}
\section{Results}
\label{sec:result}
\subsection{Mapping the projected POS magnetic fields}
The visualization of the decomposed $\rm ^{12}CO$ and $\rm ^{13}CO$ emission lines is presented in Fig.~\ref{fig:1}. There we only draw the emission line maximum of each decomposed spectrum. Compared with earlier work of \cite{2016MNRAS.457.2675H,2020NatAs...4.1064H}, in which dense tracers HNCO (J = 4-3), $\rm N_2H^+$ (J = 1-0), and HNC (J = 1-0) were adopted, $\rm ^{12}CO$ and $\rm ^{13}CO$ emissions are more diffusive and permeated in the CMZ, which allows us to trace the global magnetic fields and to compare the results with Planck polarization. We observe an apparent high-intensity filamentary structure in the velocity range of $0-100$ km/s, which was also detected by \cite{2016MNRAS.457.2675H}. This structure is believed to represent the dense gas streams in the CMZ. The figures show two distinguished structures in $-200$ - $-100$ km/s and $100-200$ km/s. These unusual high-velocity structures are most likely from a rapidly expanding ring with a radius $\sim250$ pc \citep{1972ApJ...175L.127S,1974PASJ...26..117K,2019PASJ...71S..19T}. We group the decomposed spectra into five components via the ACORNS algorithm (see Fig.~\ref{fig:2}). These five components correspond to the most significant Gaussian profile in the total spectrum averaged over the full CMZ. The components 1, 2, and 4, which have small velocities ($|v|<100$ km/s), are more associated with the gas in the galactic disk and the CMZ streams so that they may suffer more confusion in the decomposition. In contrast, components 3 and 5 come from an expanding ring showing very high velocity. Therefore they are more likely to be real physical structures.

Note that the magnetic field morphology traced by the VGT using the $\rm ^{12}CO$ and $\rm ^{13}CO$ emission lines has been presented in \citet{CMZ}. The VGT considers all contribution along the LOS and exhibits good agreement with the Planck polarization. This encourages us to decompose further the magnetic field of the high-velocity components 3 and 5 in detail. Fig.~\ref{fig:3} presents the orientation distribution of magnetic fields towards the components 3 and 5 measured by the VGT using the $\rm ^{12}CO$ and $\rm ^{13}CO$. From the decomposed intensity map, we see that both components span over the full CMZ while component 3 locates at the low latitude region ($b<0$) and component 5 focuses on the high latitude range ($b>0$). In particular, no apparent global mean magnetic field is observed at the VGT's scale $\sim5$ pc and sub-structures exhibit distinguished magnetic field orientation. This indicates a rather turbulent magnetic field picture of components 3 and 5. Also, as $\rm ^{12}CO$ and $\rm ^{13}CO$ show similar magnetic field distributions, it suggests that the magnetic field generally has only small variation over the volume density range from $10^2$ cm$^{-3}$ to $10^3$ cm$^{-3}$ around.

%The central coordinates of each sub-regions are \textbf{A}: ($l=0.63^\circ,b=-0.10^\circ$), \textbf{B:} ($l=0.05^\circ,b=0.06^\circ$), and \textbf{C:} ($l=-0.53^\circ,b=-0.04^\circ$). 
\begin{figure*}
	\includegraphics[width=.95\linewidth]{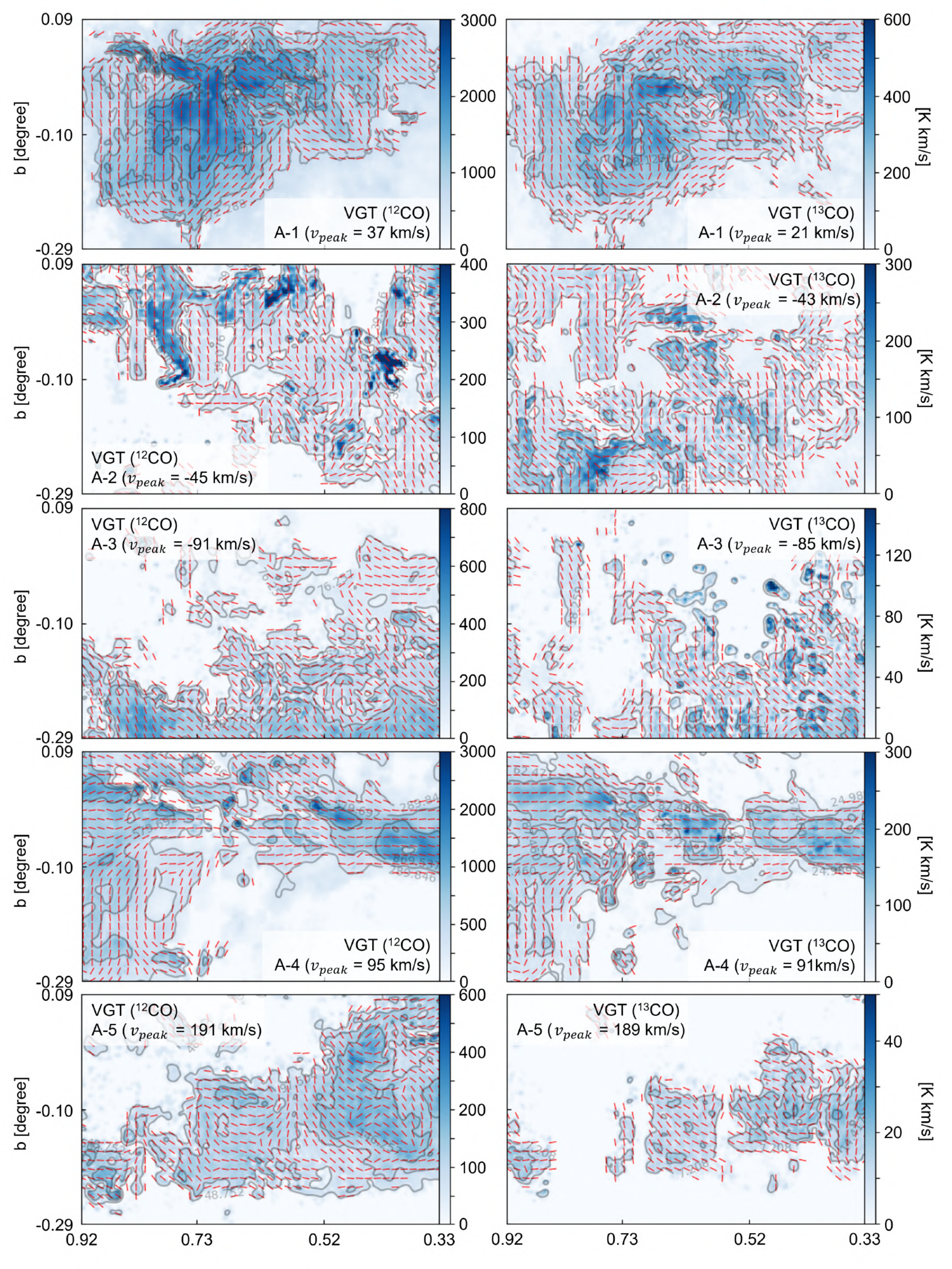}
    \caption{Visualization of the magnetic field (red segment) associated with each decomposed velocity component in the sub-region \textbf{A} (see Fig.~\ref{fig:4} and Tab.~\ref{tab:1} for details). The magnetic field is obtained from the application of the VGT to $\rm ^{12}CO$ (left) and $\rm ^{13}CO$ (right). The contours on the $\rm ^{12}CO$ intensity maps start from its median intensity.}
    \label{fig:6}
\end{figure*}

\begin{figure*}
	\includegraphics[width=0.8\linewidth]{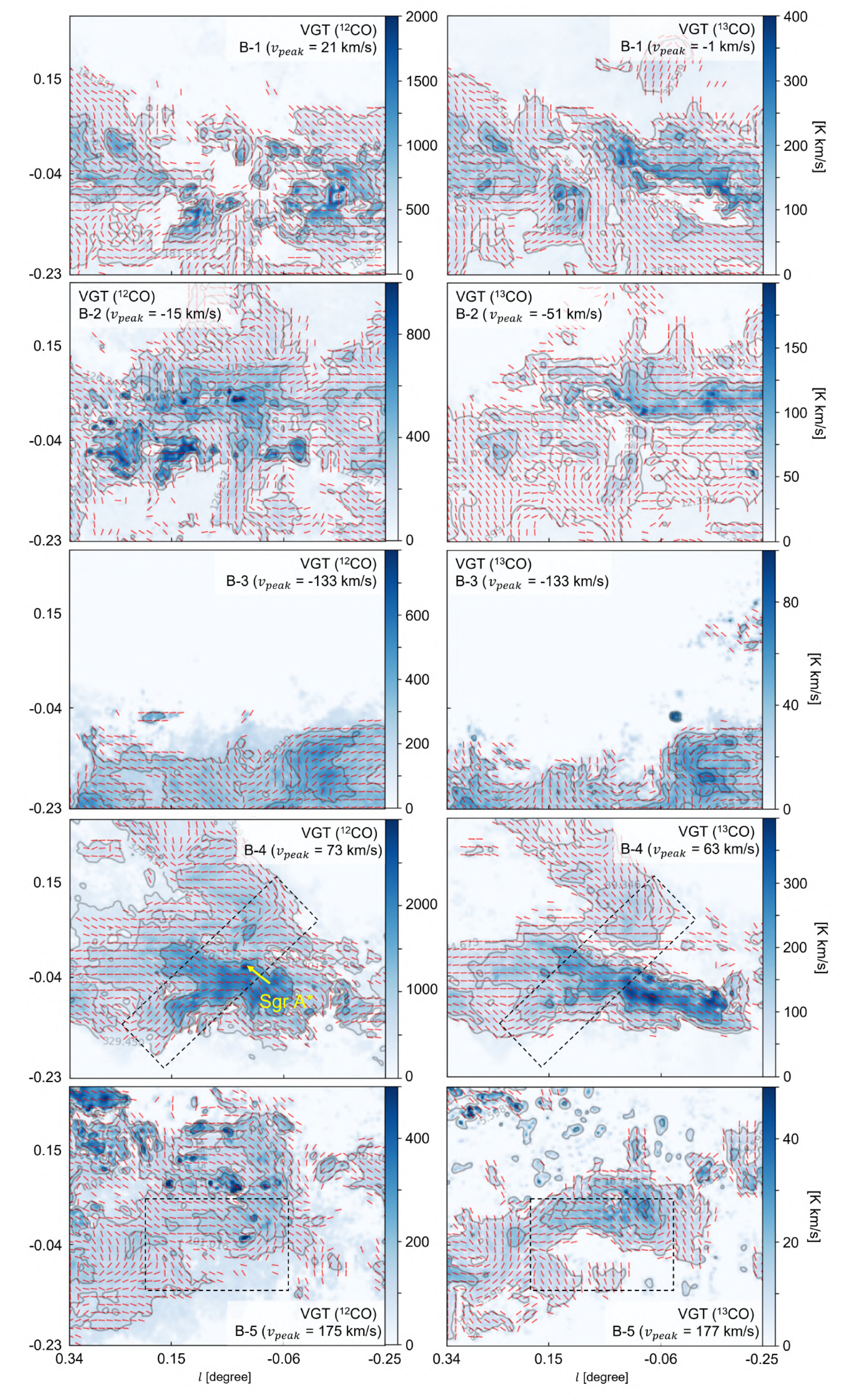}
    \caption{Same as Fig.~\ref{fig:6}, but for the sub-region \textbf{B}.}
    \label{fig:7}
\end{figure*}

\begin{figure*}
	\includegraphics[width=.95\linewidth]{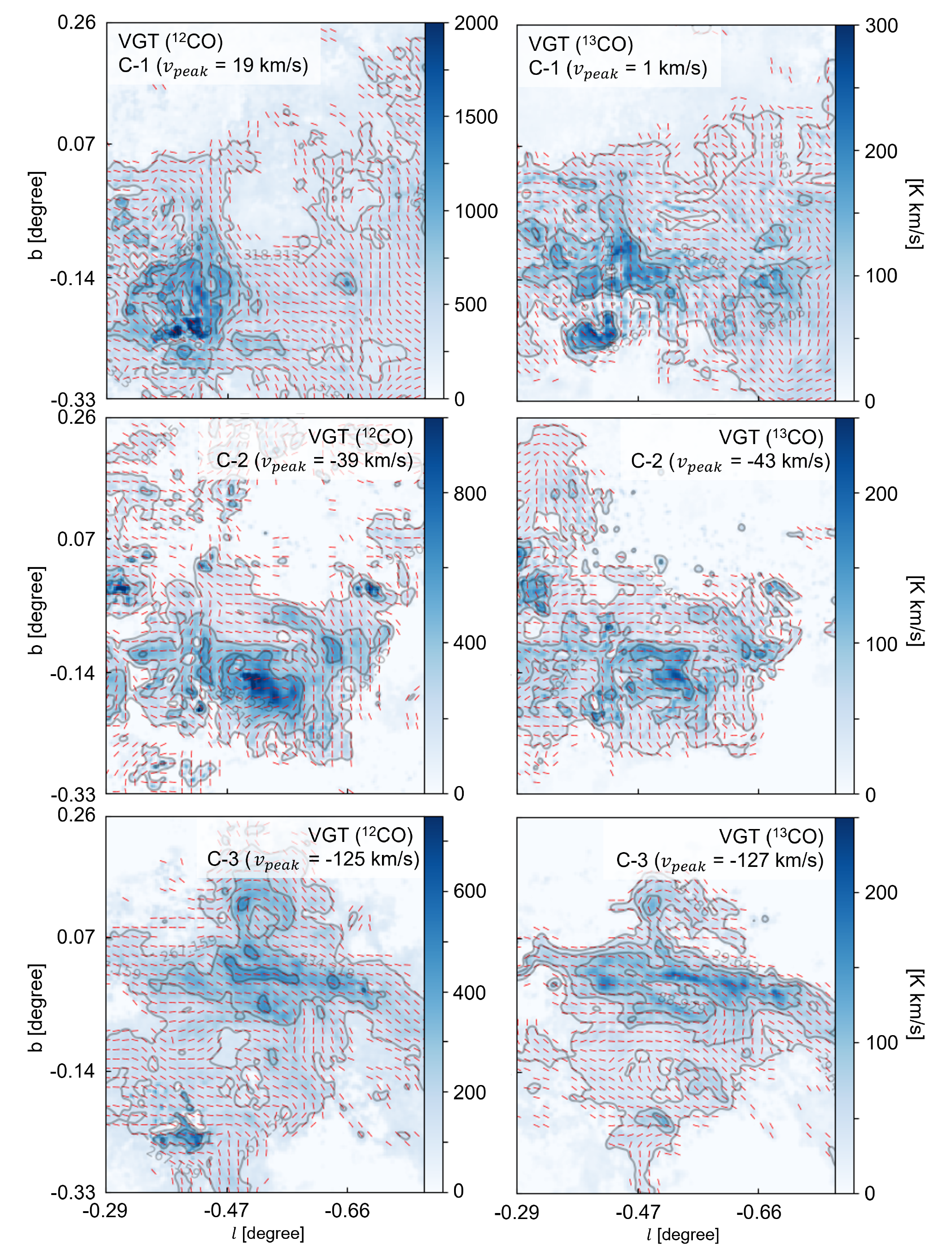}
    \caption{Same as Fig.~\ref{fig:6}, but for the sub-region \textbf{C}.}
    \label{fig:8}
\end{figure*}

\begin{figure*}
	\includegraphics[width=.95\linewidth]{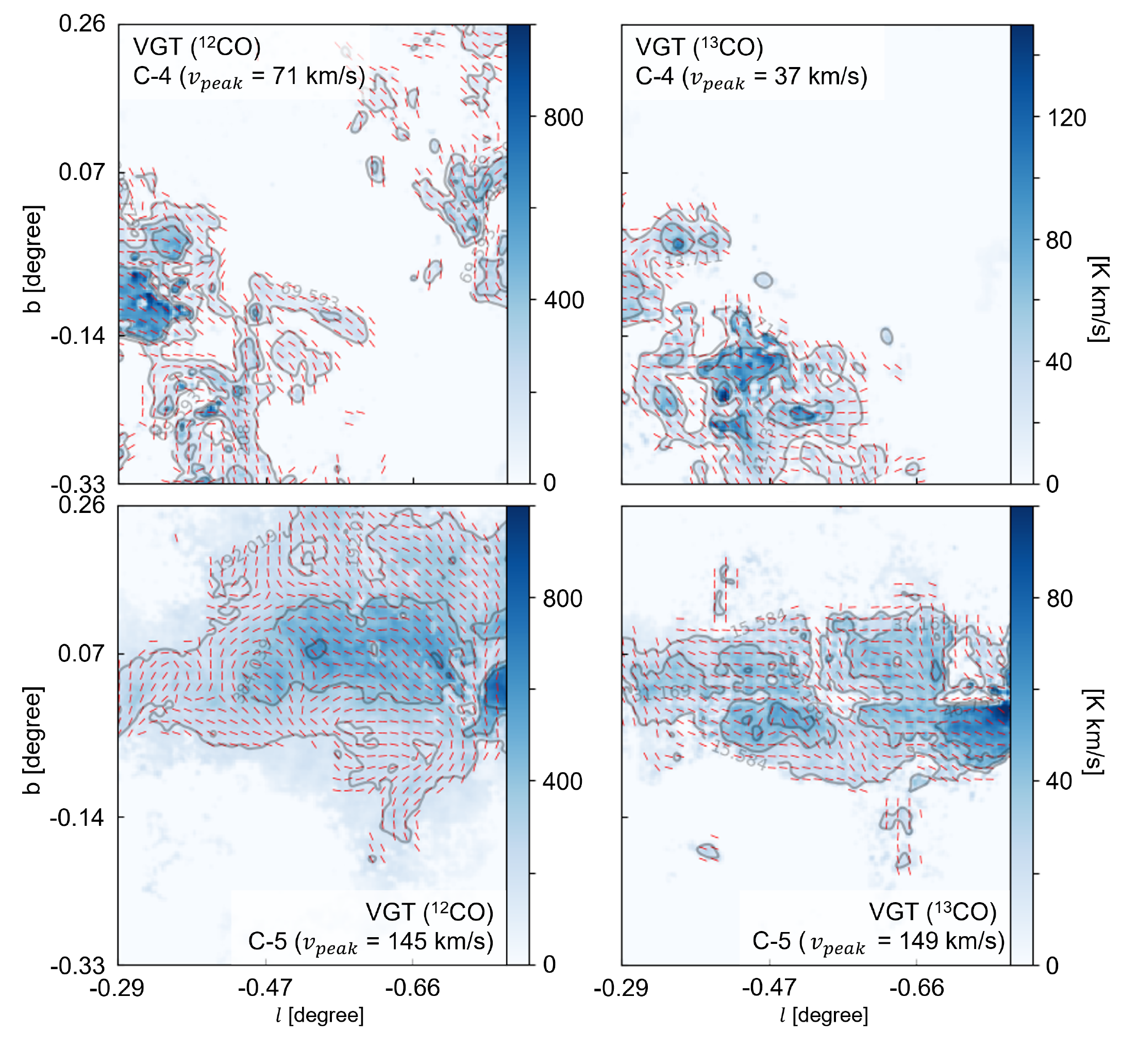}
    \caption{Same as Fig.~\ref{fig:6}, but for the sub-region \textbf{C}.}
    \label{fig:9}
\end{figure*}

\begin{figure*}
	\includegraphics[width=.95\linewidth]{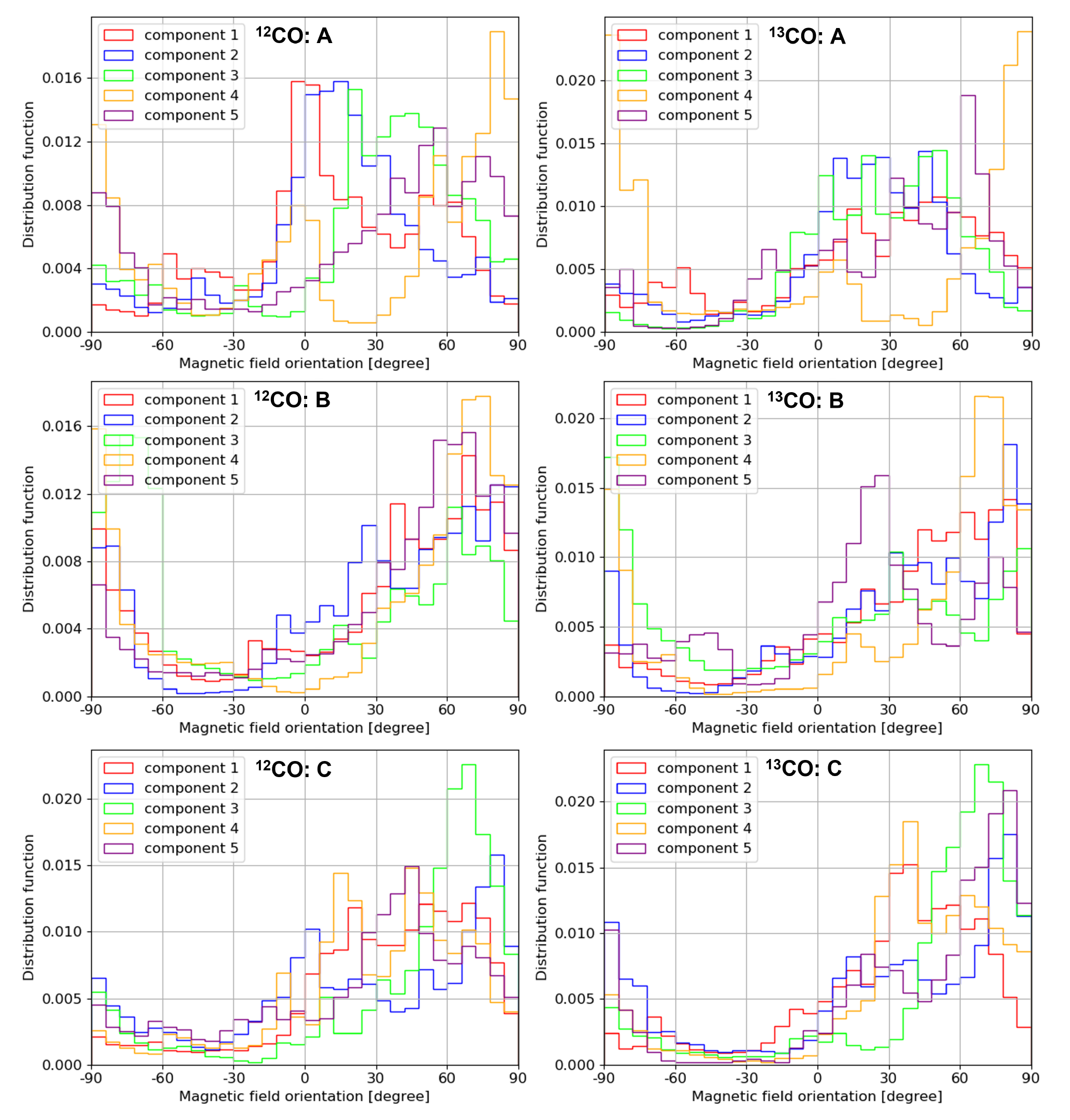}
    \caption{The histogram of the magnetic field angle obtained from VGT using $\rm ^{12}CO$ (left) and $\rm ^{13}CO$ (right). The position angle is defined as east from the Galactic north.}
    \label{fig:10}
\end{figure*}

\begin{figure}
	\includegraphics[width=.95\linewidth]{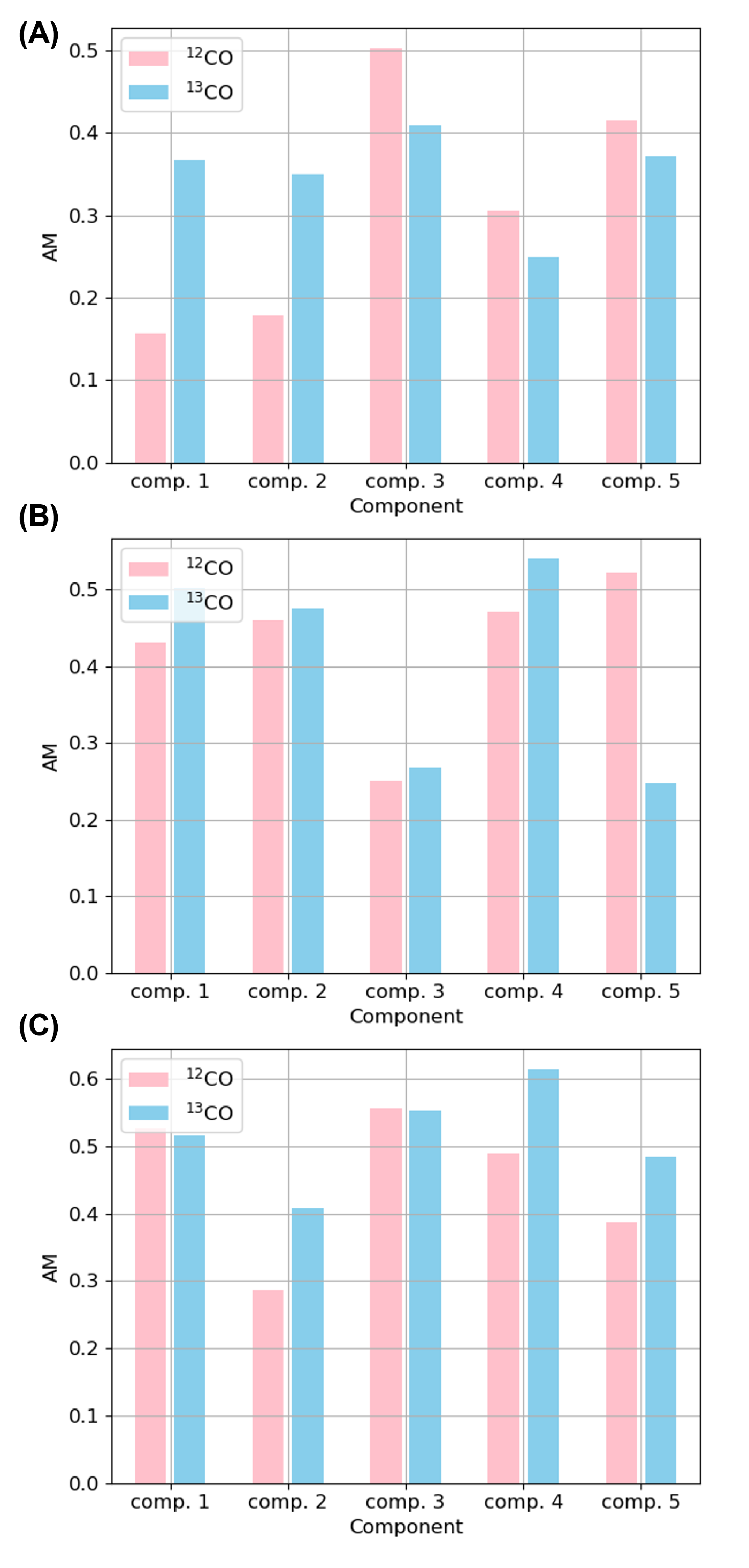}
    \caption{The AM of the magnetic fields obtained from VGT and Planck polarization. VGT is calculated from $\rm ^{12}CO$ (pink) and $\rm ^{13}CO$ (blue) for the sub-region \textbf{A} (top), \textbf{B} (middle), and \textbf{C} (bottom).}
    \label{fig:11}
\end{figure}

\subsection{Decomposition of velocity components}
\label{subsec:dec}
To reduce the confusion on low velocity components on global scales, over which a coherent structure can have a large velocity variation, we have separate close-ups of three sub-regions \textbf{A}, \textbf{B}, \textbf{C}. The sub-region \textbf{A} covers Sgr B2, similarly for other two regions \textbf{B} contains the Brick and Sgr A*, and \textbf{C} includes the Sgr C. Compared with the full CMZ, the spectra of the zoom-in regions exhibit more distinct Gaussian profiles (see Fig.~\ref{fig:4}) so that they are more likely to be real physical structures. Based on the spectra, we also decompose the sub-region into five components. 

A visualization of the decomposed velocity components is presented in Fig.~\ref{fig:4}. The spectral lines of the decomposed components are generally compatible with the un-decomposed averaged one. We observe that some lines become more significant after decomposition, such as the one of the sub-region \textbf{B} component 3 (in green color). This difference can be easily understood. The un-decomposed spectrum contains both high-intensity and low-intensity pixels, while the decomposition blanks out low-intensity pixels, in which the intensity is less than three times the RMS noise level. Without the contribution from low-intensity pixels, the averaged spectral line gets a higher intensity value after averaging. For each component, we calculate the velocity $v_{\rm peak}$ of the averaged spectral line maximum and velocity dispersion $\delta v$. The results are summarized in the Appendix Tab.~\ref{tab:1}. Apparently, for a single sub-region, the difference between $v_{\rm peak}$ is more extensive than $\delta v$. Therefore, we expected that the decomposed components are actual physical structures, and the effect of velocity caustic is minimum. 

The decomposed structures with different velocities have a more clear physical meaning. For instance the velocity component 1 (including A-1, B-1, C-1) falls in the velocity range around 0 (see Tab.~\ref{tab:1}). Although there is significant gas contamination in the Galactic disk at $v\sim0$, the bulk of the dense gas, including Sgr B2, is in the CMZ. Other velocity components with negative velocity (i.e., components 2 and 3) are mostly due to the front side of the CMZ, while positive velocity components (i.e., components 4 and 5) come from the back side. Note here that the front and backside refer to the CMZ structures on inner radius scales $\sim250$ pc. %In particular, the high velocity components 3 and 5 are most likely associated with a expanding ring around the CMZ \citep{1972ApJ...175L.127S,1974PASJ...26..117K}.

\subsection{POS magnetic fields of velocity components}
We first present the un-decomposed magnetic field maps for each sub-region compared with the Planck polarization. The results are shown in Fig.~\ref{fig:5}. For the sub-region \textbf{A}, the magnetic field orientation inferred from both $\rm ^{12}CO$ and $\rm ^{13}CO$ is along the northeast-southwest direction in low-intensity regions. However, the magnetic field inferred from $\rm ^{12}CO$ becomes vertically oriented in the high-intensity region. This change is not observed in $\rm ^{13}CO$. The gas dynamics or magnetic field properties are likely changed with respect to different volume densities. The optically thick tracer $\rm ^{12}CO$ usually samples the diffuse outskirt region with volume density $n\sim10^2$ cm$^{-3}$, while $\rm ^{13}CO$ traces the region with $n\sim10^3$ cm$^{-3}$. As for the sub-region \textbf{B}, globally magnetic fields inferred from both $\rm ^{12}CO$ and $\rm ^{13}CO$ appear northeast-southwest oriented. A change of the field orientation is observed at the superbubble around the Quintuplet cluster. The sub-region \textbf{C} shows relatively lower intensity, and the magnetic fields are along to the northeast-southwest with slight variations.

Due to the low resolution of Planck, the magnetic field's variation is slight. Therefore, we visualize the field line via the Line Integral Convolution (LIC). We find the resulting magnetic fields from $\rm ^{12}CO$ and $\rm ^{13}CO$ have good agreement with the Planck polarization showing AM from $\sim0.6$ to $\sim0.9$. However, as discussed in \cite{CMZ}, astrophysical objects may possess systematic gradients that are imposed by external conditions, which can contribute to our obtained velocity gradients. However, compared with the systematic gradients, which do not depend on scales, the amplitude of turbulence's velocity gradient increases at small scales. We expect that turbulence's velocity gradient is dominant at small scales and the contribution from other effects (such as galactic shear) to the velocity gradients is insignificant. In any case, as here we see that the velocity gradient has good agreement with Planck polarization, the contribution from other effects, which is not expected to have any correlation with the magnetic field, is in general not significant.

Next, we apply the VGT to trace the magnetic field orientation of each velocity component decomposed in \S~\ref{subsec:dec}. The magnetic fields of the sub-regions \textbf{A}, \textbf{B}, and \textbf{C} are presented in Fig.~\ref{fig:6}, Fig.~\ref{fig:7}, and Figs.~\ref{fig:8}-\ref{fig:9}, respectively. Note that low-intensity regions may include more systematic uncertainties coming from either the decomposition method or intrinsic observational error. We, therefore, blank out the magnetic field vectors where the integrated intensity is less than its median value. 

For the sub-region \textbf{A}, the dense clump ($l\sim0.73^\circ, b\sim-0.10^\circ$) seen in Fig.~\ref{fig:5} appears in component 1. The magnetic field structure is pretty complex, showing a vortex-like shape. In general, $\rm ^{12}CO$ tends to show distinctly different magnetic structures in the north-south orientations, while for $\rm ^{13}CO$, more magnetic field vectors aligned along the northeast-southwest, being similar to the magnetic field orientation observed in Fig.~\ref{fig:5}. This can be understood because component 1 has the most significant intensity (see Fig.~\ref{fig:4}) so that its contribution dominates the projected magnetic field map, which is contributed by all components. In addition, the upper left filamentary structure in \textbf{A-1} may be due to the collapse of gas along the vertical magnetic field, consistent with the theoretical prediction. As for components 2 and 3, the observed intensity structures are quite different in $\rm ^{12}CO$ and $\rm ^{13}CO$, although the magnetic field is globally-oriented along the northeast-southwest direction. The different structures may be due to the complexity of CMZ's streams in the intermediate velocity range, so more confusion exists on the LOS. Component 4 has the second most significant intensity appearing as a filamentary structure. Its magnetic field orientation is preferentially along the horizontal (west-east) direction, which also appears in the projected magnetic field map (see Fig.~\ref{fig:5}). A toroidal field dominates this magnetic field. Also, we plot the histogram of magnetic field orientation for each component in Fig.~\ref{fig:10}. We can see that for both $\rm ^{12}CO$ and $\rm ^{13}CO$, other less significant components 2, 3, 5 have the majority of magnetic field angle is in the range of $0 - 60^\circ$ (east from the north).

As for the sub-region \textbf{B}, components 1, 2, and 4 dominate the observed molecular emission. Similar to the sub-region \textbf{A}, component 2 with negative moderate velocity exhibits quite different intensity structures in two CO emissions. Magnetic fields in the central area of this sub-region are mainly built upon components 1, 2, and 4. Two preferential magnetic field directions are oriented along the northeast-southwest or the east-west. This trend is more clear in the histogram (see Fig.~\ref{fig:10}), which shows that the magnetic field is mostly oriented in the position angle range of $30^\circ - 60^\circ$ (east from the north). However, several small clumps show a change of magnetic field direction. Two locate around $l\sim0.15^\circ, b\sim-0.10^\circ$ in components 1, 4, and 5 and $l\sim-0.10^\circ, b\sim0.00^\circ$ in component 1. Another is around $l\sim-0.06^\circ, b\sim-0.10^\circ$ in component 3. There the magnetic fields change the direction being along the northwest-southeast. The change of direction is also observed in the un-decomposed magnetic field map (see Fig.~\ref{fig:1}) in the same location as components 1, 4, and 5. The change seen in component 1 may come from the superbubble around the Quintuplet cluster. A significant hydro effect could change the direction of the velocity gradient. As for component 4, which is close to Sgr A*, the effect of outflows perpendicular to Sgr A* streams could change the direction of magnetic fields (see the dashed box in Fig.~\ref{fig:7}). Components 3 and 5 show very high velocity but were rarely imaged before. %The component 5 appears as a circular shape there, which is more apparent in $\rm ^{13}CO$. The change of magnetic field there is associated with the expanding rings.

The sub-region \textbf{C} is less complicated, although several whirling structures appear there in particular in components 2, 3, and 5. Component 1 has the most significant intensity, while other components are at a similar level (see Fig.~\ref{fig:2}). There are three preferential magnetic field directions along the northeast-southwest, north-south, or east-west. The latter two suggest the dominance of poloidal and toroidal fields, respectively. As shown in Fig.~\ref{fig:10}, the magnetic field is mainly orientated in the range of $0^\circ - 90^\circ$ (east from the north) with a peak value around 45$^\circ$ or 70$^\circ$, which could come from the superposition of poloidal and toroidal fields.

In Fig.~\ref{fig:11}, we compute the AM of the VGT measurement and Planck polarization for each component. A higher value of the AM suggests that the corresponding component contributes more to the projected polarization measurement. As we see, for the sub-region \textbf{A}, both $\rm ^{12}CO$ and $\rm ^{13}CO$ of components 3, 4, and 5 give similar AM values, which suggest the magnetic field's variation is relatively small at two density ranges. However, a significant difference appears in components 1 and 2, in which $\rm ^{12}CO$'s AM drastically drops. As we discussed above, the difference comes from the fact that the magnetic field inferred from $\rm ^{12}CO$ changes its direction to be along the north-south. The highest intensity value of component 1 ensures this change is observable in an un-decomposed magnetic field map. This change is not observed in Planck polarization and VGT-$\rm ^{13}CO$ measurement. It is likely that the polarization is dominated by the contribution from denser regions.

The situation is similar for the sub-region \textbf{B}. For both tracers, components 1 - 4 have similar AM. The AM of component 3 is relatively low. Considering that the magnetic field inferred from the Planck polarization is mainly along the northwest-southeast, this low AM is contributed by the magnetic fields oriented along the northwest-southeast and east-west. Also, the $\rm ^{13}CO$'s AM of component 5 significantly drops, while $\rm ^{12}CO$ does not. The $\rm ^{13}CO$ case is due to the magnetic field associated with the observed circular structure (see Fig.~\ref{fig:7}) is swirling, but the Planck polarization gives a northeastern magnetic field direction. This low AM associated with the swirling structure is compensated by the northeastern structure that appeared in $\rm ^{12}CO$.

As for the sub-region \textbf{C}, the AM of $\rm ^{12}CO$ and $\rm ^{13}CO$ has similar value. As their spectral intensities are at the same level, all components have approximately equal contributions to the total magnetic field map.

\begin{figure}
	\includegraphics[width=1.0\linewidth]{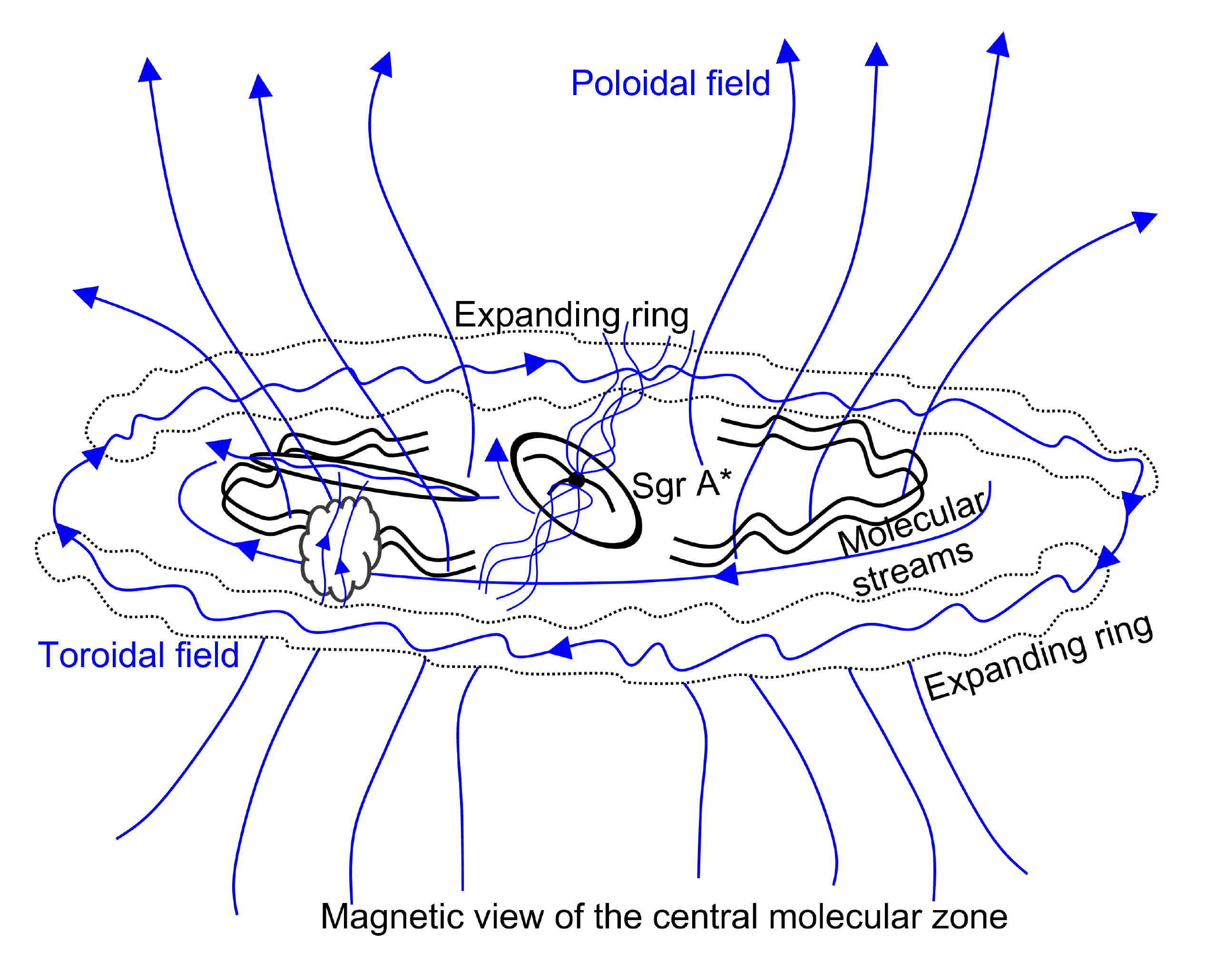}
    \caption{Cartoon of the magnetic field configuration in the CMZ. Dotted lines indicate relatively lower density.}
    \label{fig:12}
\end{figure}

\section{Discussion}
\label{sec:dis}
\subsection{Magnetic field tomography with VGT}
Dust polarization is an effective way of tracing integrated magnetic fields along the LOS. It, however, meets difficulties in separating the magnetic field associated with different components of a single cloud or various clouds along the same LOS.

The VGT provides a solution to the difficulties. This work shows that emissions of different velocity components or clouds along the same LOS can be decomposed in the position-position-velocity space. The VGT can then trace the magnetic fields associated with individual components/clouds, resulting in a magnetic field tomography. Here we use the CMZ as an example, while the application to the supernova remnant W44 is presented in \citealt{Liu21}. When summing up contributions from all components, we show that the VGT-inferred results globally agree with those from the Planck dust polarization measurements. This agreement increases our confidence in decomposing the magnetic fields. On the other side, the decomposition approach here benefits the polarization study in distinguishing the contribution from different regions along the LOS. For instance, our study shows that several components can dominate the agreement with Planck polarization. As shown in Fig.~\ref{fig:11}, the best agreements appear in components A-3 and A-5. It suggests these components mainly contribute to the signal of polarized dust emission. This opens a new way to study dust's properties in the galactic center. Note here we separately decompose $\rm ^{12}CO$ and $\rm ^{13}CO$. The decomposition may contain intrinsic uncertainty, and we view our present study as the first step in this direction.  
%so that the irrespective components can be significantly different. A joint decomposition, which may be developed further, can help to reduce thos uncertainty.

The VGT approach of decomposing magnetic fields introduced here also works for other objects on smaller scales, such as the Sgr A west region and the Brick, which several molecular species can trace. The VGT approach will allow us to access the magnetic field tomography along the LOS. Also, different tracers, which sample various gas density ranges or phases, will enable us to explore the magnetic fields associated with density or temperature ranges. For instance, the VGT using $\rm ^{12}CO$ reveals the magnetic field associated with diffuse molecular gas with volume density $\sim10^{2}$ cm$^{-3}$, while $\rm ^{13}CO$ can give information on higher volume density $\sim10^{3}$ cm$^{-3}$. This decomposition method is also expected to work for other magnetic field tracing methods based on MHD turbulence's anisotropy. The methods include the Correlation Function Analysis (CFA; \citealt{2011ApJ...740..117E}) and the Structure-Function Analysis (SFA; \citealt{2021ApJ...911...37H,2021ApJ...910...88X,2021ApJ...915...67H}).

\subsection{Magnetic field configuration in the Galactic center}
Owing to the advantages of the VGT, one can further explore the magnetic field configuration in the Galactic center. Firstly, the magnetic fields preferentially orientate along the northeast/southwest direction. This orientation may come from the superposition of a toroidal field and a poloidal field. In particular, we observe apparent vertical magnetic field components in the sub-regions A-1 (around $l\sim0.73^\circ,b\sim-0.10^\circ$) and C-1 (around $l\sim-0.45^\circ,b\sim-0.14^\circ$). Since velocity components, A-1 and C-1 with velocity concentrate on 0 km/s (see Tab.~\ref{tab:1} and Fig.~\ref{fig:3}) are mainly from the Galactic disk and the streams associated with the CMZ. This suggests that a dominated role of poloidal fields. 

It is natural to assume that this poloidal magnetic field is generated by the galactic dynamo \citep{1999ARA&A..37...37K}. It is widely believed that the process amplifies the magnetic field starting with a seed magnetic field with differential rotation of the galaxy naturally producing a toroidal field through magnetic field stretching \citep{1992A&A...264..326L}. This is often termed the $\Omega$-effect. The classical galactic dynamo assumes a cyclonic motion of turbulence can make a poloidal loop from the toroidal field through the so-called $\alpha$-effect \citep{1988ASSL..133.....R}. The differential rotation can shear the newly formed loop further to amplify the toroidal component of the magnetic field. Different from the galactic disk, the poloidal field near a spherical object, i.e., the galactic CMZ, is more likely to be dipole and does not change its direction \citep{1979cmft.book.....P}. An additional effect that can be important includes gas compression that amplifies the vertical field (with respect to galactic disk) in the galactic center, while ambipolar diffusion removes the parallel component \citep{2000ApJ...528..723C}.

The problem of the picture above is that the classical dynamo \citep{1971ApJ...163..255P} does not conserve magnetic helicity, which is a necessary constraint for a dynamo in a highly conducting medium \citep{1994PhRvL..72.1651G}. At the same time, the model that accounts for the magnetic helicity conservation \citep{2001ApJ...550..752V,2014ApJ...780..144V} does not have such a simple topological explanation. Similarly, a less obvious process of magnetic reconnection in turbulent medium induces the process of reconnection diffusion \citep{LV99,2021MNRAS.503.1290S} that is expected to move the magnetic field in turbulent gas faster than ambipolar diffusion \citep{2015ASSL..407..311L}. A further discussion of these theoretical issues is well beyond the scope of the present paper. However, we believe that the magnetic field structure of the CMZ revealed in the present paper can help constrain future theoretical constructions.   

Several horizontal magnetic fields are also observed in the components such as A-4 and B-4. The toroidal magnetic field could play a crucial role in these regions. In the region around the Sgr A* (i.e., B-4), the stream or other physical effects could be sufficiently strong to stretch the magnetic fields. However, a change of the magnetic field orientation around 90 degrees is found on the upper and lower part of the cloud around the Sgr A* (see the dashed box in Fig.~\ref{fig:7}). This apparent change results from an outflow perpendicular to the Sgr A* streams. 

A critical aspect of the present work is the decomposition of the high velocity expanding ring (i.e., components 3 and 5). Its distinguished high velocity ensures minimum confusion (i.e., some clouds with the same velocity at different spatial positions along the LOS may be identified as one component) along the LOS. A distinct swirling magnetic field structure is observed in component B-5 (velocity $\sim175$ km/s). In particular, the expanding ring appears with no apparent mean magnetic field, suggesting a highly turbulent field there. Based on the decomposed magnetic fields in this work, we present a simple cartoon of the magnetic field configuration in the CMZ, as shown in Fig.~\ref{fig:12}.
%On the other side, the decomposition paradigm introduced here benefits the polarization study in distinguishing the contribution from different regions along the LOS. For instance, as shown in Fig.~\ref{fig:11}, polarization exhibits best agreement with VGT in components A-3 and A-5. It suggests the contribution for the foreground and background is more significant in the sub-region \textbf{A}. However, for the sub-region \textbf{B} and \textbf{C}, component 1 appears better agreement. It indicates the significance of the Galactic disk's contribution to polarization.

\section{Summary}
\label{sec:con}
This study uses the Velocity Gradient Technique (VGT) to trace the magnetic fields and disentangle the confusion of magnetic fields associated with different LOS velocity structures in the CMZ region. The main results that we obtained are summarized as follows:
\begin{enumerate}
    \item Using the SCOUSEPY algorithm, we decompose five velocity components at different systemic velocities in the diffuse $\rm^{12}CO$ and moderately dense $\rm ^{13}CO$ gas. The combination of these two molecular tracers allows us to probe gas in the CMZ over an extensive opacity range. The decomposition is performed in the entire CMZ region and three zoom-in sub-regions.
    \item We find two filamentary $\rm ^{12}CO$ and $\rm ^{13}CO$ structures in the high velocity range of $\pm (100 - 200)$ km/s. These structures are associated with an expanding ring in the CMZ, as proposed by \citet{1972ApJ...175L.127S} and \cite{1974PASJ...26..117K}.
    % \item We introduce a paradigm for decomposing magnetic fields in position-position-velocity space along the line-of-sight with VGT.
    \item We show that the projected magnetic fields towards the CMZ traced by the VGT are globally compatible with the Planck polarization. In general, 
    magnetic field orientations inferred from two CO isotopologs are broadly consistent, expect for a few distinct dense regions, such as the Sgr B2 and Sgr A*
    \item We present the decomposed magnetic field maps of the expanding ring and three sub-regions in the CMZ and investigate their significance. In particular, we observe a nearly vertical magnetic field orientation in the dense clump near the Sgr B2 and a change of magnetic field in the outflow regions around Sgr A*. Several structures, especially the expanding ring, show distinct swirling magnetic field orientation. 
    \item We present a schematic magnetic field configuration in the Galactic center based on the decomposed magnetic fields. Applying the VGT to denser tracers or comparing with higher resolution polarization measurements can further complete this configuration.
\end{enumerate}

\section*{Acknowledgements}
Y.H. acknowledges the support of the NASA TCAN 144AAG1967. A.L. acknowledges the support of the NSF grant AST 1715754 and NASA ATP AAH7546. We acknowledge the Nobeyama Observatory for providing the data of the CMZ region.

%%%%%%%%%%%%%%%%%%%%%%%%%%%%%%%%%%%%%%%%%%%%%%%%%%
\section*{Data Availability}
The data underlying this article will be shared on reasonable
request to the corresponding author.



\bibliographystyle{mnras}
\bibliography{reference} % if your bibtex file is called example.bib

%\begin{thebibliography}{99}
%\end{thebibliography}

%%%%%%%%%%%%%%%%%%%%%%%%%%%%%%%%%%%%%%%%%%%%%%%%%%

%%%%%%%%%%%%%%%%% APPENDICES %%%%%%%%%%%%%%%%%%%%%

\appendix

\section{Summary of the decomposed emission lines}
We use the ACORNS clustering to group the decomposed spectrum for the entire CMZ. Tab.~\ref{tab:1} presents the details of each decomposed component in the sub-regions \textbf{A}, \textbf{B}, and \textbf{C}. 
\begin{table*}
	\centering
	\caption{Parameters of the decomposed components for each sub-region (see Fig.~\ref{fig:1}), i.e., \textbf{A}, \textbf{B}, and \textbf{C}. For each component, $v_{\rm peak}$ is the velocity of the sub-region-averaged emission line maximum (i.e., $T_{\rm mb}^{\rm peak}$) and $\delta v$ is velocity dispersion calculated from normalized velocity centroid (i.e., moment-1 map). The number of decomposed spectrum (i.e., Gaussian component) is given.}
	\label{tab:1}
	\begin{tabular}{l  c  c  c  c  c  c} % four columns, alignment for each
		\hline
		Emission & Region & Component & $v_{\rm peak}$ [km s$^{-1}$] & $\delta v$ [km s$^{-1}$] & $T_{\rm mb}^{\rm peak}$ [K] & Number of spectra\\
		\hline\hline
		\multirow{15}{*}{$\rm ^{12}CO$} & \multirow{5}{*}{\textbf{A}} & 1 & 37.0 & 19.64 & 18.60 & 63856\\
		& & 2 & -45.0 & 12.43 & 5.22 & 29609\\
		& & 3 & -91.0 & 17.04 & 4.85 & 41473\\
		& & 4 & 95.0 & 17.03 & 16.45 & 46749\\
		& & 5 & 191.0 & 7.57 & 6.52 & 30953\\\cline{2-7}
		& \multirow{5}{*}{\textbf{B}} & 1 & 21.0 & 16.39 & 12.44 & 54272\\
		& & 2 & -15.0 & 20.86 & 5.06 & 94462\\
		& & 3 & -133.0 & 13.85 & 7.25 & 38550\\
		& & 4 & 73.0 & 26.29 & 14.36 & 53245\\
		& & 5 & 175.0 & 14.26 & 5.11 & 48586\\\cline{2-7}
		& \multirow{5}{*}{\textbf{C}} & 1 & 19.0 & 16.06 & 8.39 & 101502\\
		& & 2 & -39.0 & 18.11 & 12.49 & 53387\\
		& & 3 & -125.0 & 21.25 & 7.33 & 71216\\
		& & 4 & 71.0 & 24.21 & 5.89 & 19230\\
		& & 5 & 145.0 & 12.41 & 7.05 & 61076\\
		\hline
		\multirow{15}{*}{$\rm ^{13}CO$} & \multirow{5}{*}{\textbf{A}} & 1 & 21.0 & 16.30 & 3.07 & 80639\\
		& & 2 & -43.0 & 19.96 & 1.69 & 60412\\
		& & 3 & -85.0 & 29.75 & 0.76 & 24770\\
		& & 4 & 91.0 & 13.93 & 2.61 & 38740\\
		& & 5 & 189.0 & 8.90 & 0.89 & 13886 \\\cline{2-7}
		& \multirow{5}{*}{\textbf{B}} & 1 & -1.0 & 12.82 & 2.41 & 102575\\
		& & 2 & -51.0 & 12.05 & 1.61 & 57204\\
		& & 3 & -133.0 & 21.33 & 1.13 & 23930\\
		& & 4 & 63.0 & 25.80 & 2.20 & 44418\\
		& & 5 & 177.0 & 16.00 & 0.82 & 24832\\\cline{2-7}
		& \multirow{5}{*}{\textbf{C}} & 1 & 1.0 & 10.12 & 2.68 & 91397\\
		& & 2 & -43.0 & 17.96 & 1.73 & 44848\\
		& & 3 & -127.0 & 18.24 & 1.34 & 43570\\
		& & 4 & 37.0 & 17.34 & 1.00 & 17780\\
		& & 5 & 149.0 & 12.43 & 1.05 & 26243\\
		\hline
	\end{tabular}
\end{table*}

%If you want to present additional material which would interrupt the flow of the main paper, it can be placed in an Appendix which appears after the list of references.

%%%%%%%%%%%%%%%%%%%%%%%%%%%%%%%%%%%%%%%%%%%%%%%%%%

% Don't change these lines
\bsp	% typesetting comment
\label{lastpage}
\end{document}